\documentclass[12pt]{article}
\usepackage{html}
\usepackage{epsf}
\usepackage{psfig}
\usepackage{graphicx}
\usepackage{amssymb}
\begin{document}
\title{MATTERS OF GRAVITY, The newsletter of the APS Topical Group on 
Gravitation}
\begin{center}
{ \Large {\bf MATTERS OF GRAVITY}}\\ 
\bigskip
\hrule
\medskip
{The newsletter of the Topical Group on Gravitation of the American Physical 
Society}\\
\medskip
{\bf Number 32 \hfill Fall 2008}
\end{center}
\begin{flushleft}
\tableofcontents
\vfill\eject
\section*{\noindent  Editor\hfill}
David Garfinkle\\
\smallskip
Department of Physics
Oakland University
Rochester, MI 48309\\
Phone: (248) 370-3411\\
Internet: 
\htmladdnormallink{\protect {\tt{garfinkl-at-oakland.edu}}}
{mailto:garfinkl@oakland.edu}\\
WWW: \htmladdnormallink
{\protect {\tt{http://www.oakland.edu/physics/physics\textunderscore people/faculty/Garfinkle.htm}}}
{http://www.oakland.edu/physics/physics_people/faculty/Garfinkle.htm}\\

\section*{\noindent  Associate Editor\hfill}
Greg Comer\\
\smallskip
Department of Physics and Center for Fluids at All Scales,\\
St. Louis University,
St. Louis, MO 63103\\
Phone: (314) 977-8432\\
Internet:
\htmladdnormallink{\protect {\tt{comergl-at-slu.edu}}}
{mailto:comergl@slu.edu}\\
WWW: \htmladdnormallink{\protect {\tt{http://www.slu.edu/colleges/AS/physics/profs/comer.html}}}
{http://www.slu.edu//colleges/AS/physics/profs/comer.html}\\
\bigskip
\hfill ISSN: 1527-3431

\begin{rawhtml}
<P>
<BR><HR><P>
\end{rawhtml}
\end{flushleft}
\pagebreak
\section*{Editorial}

The next newsletter is due February 1st.  This and all subsequent
issues will be available on the web at
\htmladdnormallink 
{\protect {\tt {http://www.oakland.edu/physics/Gravity.htm}}}
{http://www.oakland.edu/physics/Gravity.htm} 
All issues before number {\bf 28} are available at
\htmladdnormallink {\protect {\tt {http://www.phys.lsu.edu/mog}}}
{http://www.phys.lsu.edu/mog}

Any ideas for topics
that should be covered by the newsletter, should be emailed to me, or 
Greg Comer, or
the relevant correspondent.  Any comments/questions/complaints
about the newsletter should be emailed to me.

A hardcopy of the newsletter is distributed free of charge to the
members of the APS Topical Group on Gravitation upon request (the
default distribution form is via the web) to the secretary of the
Topical Group.  It is considered a lack of etiquette to ask me to mail
you hard copies of the newsletter unless you have exhausted all your
resources to get your copy otherwise.

\hfill David Garfinkle 

\bigbreak

\vspace{-0.8cm}
\parskip=0pt
\section*{Correspondents of Matters of Gravity}
\begin{itemize}
\setlength{\itemsep}{-5pt}
\setlength{\parsep}{0pt}
\item John Friedman and Kip Thorne: Relativistic Astrophysics,
\item Bei-Lok Hu: Quantum Cosmology and Related Topics
\item Gary Horowitz: Interface with Mathematical High Energy Physics and
String Theory
\item Beverly Berger: News from NSF
\item Richard Matzner: Numerical Relativity
\item Abhay Ashtekar and Ted Newman: Mathematical Relativity
\item Bernie Schutz: News From Europe
\item Lee Smolin: Quantum Gravity
\item Cliff Will: Confrontation of Theory with Experiment
\item Peter Bender: Space Experiments
\item Jens Gundlach: Laboratory Experiments
\item Warren Johnson: Resonant Mass Gravitational Wave Detectors
\item David Shoemaker: LIGO Project
\item Stan Whitcomb: Gravitational Wave detection
\item Peter Saulson and Jorge Pullin: former editors, correspondents at large.
\end{itemize}
\section*{Topical Group in Gravitation (GGR) Authorities}
Chair: David Garfinkle; Chair-Elect: 
Stan Whitcomb; Vice-Chair: Steve Detweiler. 
Secretary-Treasurer: Gabriela Gonzalez; Past Chair:  Dieter Brill;
Delegates:
Alessandra Buonanno, Bob Wagoner,
Lee Lindblom, Eric Poisson,
Frans Pretorius, Larry Ford.
\parskip=10pt

\vfill
\eject

\section*{\centerline
{Remembering Wheeler}}
\addtocontents{toc}{\protect\medskip}
\addtocontents{toc}{\bf GGR News:}
\addcontentsline{toc}{subsubsection}{
\it Rembembering Wheeler, by Jim Isenberg}
\parskip=3pt
\begin{center}
Jim Isenberg, University of Oregon
\htmladdnormallink{isenberg-at-uoregon.edu}
{mailto:isenberg@uoregon.edu}
\end{center}

John Wheeler died a few months ago, at the age of 96. His life spanned a 
revolutionary era in physics which saw all of our ideas on space, time, and 
the universe completely reworked. Wheeler's ideas played a major role in all 
of this. Much more important, however, is the effect Wheeler had through his 
students. The students he inspired and cajoled and loved have been leaders of 
the physics revolution from the 1940's to this day.

The first time I really talked to John Wheeler was during the student strike 
of the late spring of 1970. I was a freshman at Princeton, going door to door 
for something called the Movement for a New Congress. I hated it. Usually, no 
one came to the door; when they did, they usually weren't interested in my 
spiel: `Are you willing to support our efforts to elect congressmen who want 
to stop the invasion of Cambodia and end the war in Vietnam?'

One day, when I was close to quitting, I saw that the next house was John 
Wheeler's. I knew Wheeler was famous for having done important work on the 
hydrogen bomb; he was also known as a strong opponent of the strike. I 
didn't expect much.

I hit the jackpot: not for the Movement, but for me. Wheeler listened 
politely, and then said, `Let me tell you about how photons can orbit around 
a black hole.'  I wasn't sure what a black hole was, nor was I sure if he knew 
I was a physics student. It didn't matter. After two hours, I was convinced 
that black holes were the best things to be found in the universe, and I was 
ready to dive into one just to see what those orbiting photons would look 
like from the inside.

General relativity was proposed by Einstein in 1915, and became very popular 
with the public in 1919, but for decades it was viewed by most physicists as 
a scientific dead end. The ideas of curved space and time were very appealing, 
but it was not at all clear that the theory's predictions were testable by 
experiment.  It was primarily Wheeler (together with Dennis Sciama in Britain, 
and Peter Bergmann in New York) who during the 1960's revived general 
relativity as a vital science, with important things to tell us about 
cosmology and astrophysics. Wheeler was well known for his work on nuclear 
fission and fusion and the brilliant discoveries made by his student Richard 
Feynman, so when Wheeler and his students began to talk seriously about 
general relativity again, physicists paid attention. They found that not only 
do such fascinating ideas as black holes, wormholes and gravitational waves 
arise from the theory, but that these ideas might lead to an understanding 
of real, observable, and testable physical effects that profoundly shape 
the universe.

Despite its reputation for innovation, in many ways physics is a conservative 
enterprise. To convince people to take general relativity seriously, Wheeler 
first needed to train a cadre of brilliant students and postdocs who could 
make important discoveries using relativity. He did this: Some, like Kip 
Thorne and Jacob Bekenstein, focused on black holes and astrophysics; others, 
like Charlie Misner, Dieter Brill, Bob Geroch, Demetrios Christodoulou and 
Jimmy York, worked on the mathematics of general relativity; still others, 
like Bob Wald, Bill Unruh and Hugh Everett, thought hard about the relation 
between quantum theory and general relativity. Together they produced a sea 
change in how we think about the universe.

I didn't work with Wheeler directly until almost two years after that first 
encounter. But I did learn a huge amount from him the next year, by wandering 
around Jadwin Hall (home of Princeton's physics department) at night. Often, 
peeking into darkened lecture halls, you could still see the physics lecture 
masterwork of Wheeler on multi-storied blackboards covered with beautifully 
colorful representations of black holes, wormholes and expanding universes, 
together with the equations which modeled these things. Even before I knew 
much general relativity, I loved trying to work through those lectures, 
all from the blackboard drawings. 

When I started working with Wheeler in my junior year, I learned a huge 
amount and not just physics. Wheeler understood that scientific discovery is 
not enough. You need to be able to tell people about it effectively and 
convincingly. So, a key feature of every Wheeler lecture was that you 
remembered something interesting, something he wanted you to remember. It 
was no accident that Wheeler coined the term `black hole' for those strange 
collapsed stars which he and others were just beginning to understand during 
the 1960s. These theoretical objects had earlier been called `frozen stars,' 
but Wheeler knew that a frozen star is chilling and forgettable while a black 
hole is awesome and fascinating.

Wheeler's writing was just as memorable as his lectures. Some have found his 
writing, sprinkled as it is with phrases like `it from bit' and `mass without 
mass,' a bit strange. This point was brought home to me when I submitted my 
first scientific article to a leading physics journal. Wheeler had gone over 
the article with me many times, and had suggested a multitude of changes, 
which I dutifully made. The referee report was short. It said that the science 
seemed fine, but `Wheeler can write like this; you can't!'

I still don't write like Wheeler. However, as I and other physicists await 
the word from the Laser Interferometer Gravitational-Wave Observatory that we 
can detect gravitational radiation, and as we discuss how we can use these 
observations to probe the physics of black holes (without having to dive into 
them ourselves), I think it a worthy goal to strive to do science like Wheeler.

----------------------------------------------------

This article, with a few small changes, was published last May in The Register Guard, which is the Eugene, Oregon daily newspaper.

\vfill\eject
\section*{\centerline
{A Brief Summary of the WMAP5 Results}}
\addtocontents{toc}{\protect\medskip}
\addtocontents{toc}{\bf Research Briefs:}
\addcontentsline{toc}{subsubsection}{
\it A Brief Summary of the WMAP5 Results, by Lyman Page}
\parskip=3pt
\begin{center}
Lyman Page, Princeton University
\htmladdnormallink{page-at-princeton.edu}
{mailto:page@princeton.edu}
\end{center}
  
\newcommand{\wmap}    {{\sl WMAP}}
\newcommand{\map}    {{\sl WMAP}}
\newcommand{\cobe}   {{\sl COBE}}
\newcommand{\etal}   {{\it et al.}}
\newcommand{\n}      {{\bf{n}}}
\newcommand{\EV}[1]  {\langle#1\rangle}
\newcommand{\lmax}   {l_{\rm max}}
\newcommand{\mmax}   {m_{\rm max}}
\newcommand{\uKsq}   {\mbox{$\mu{\rm K}^2$}}
\newcommand{\dg}     {\mbox{$^{\circ}$}}
\newcommand{\lsim}   {\mbox{$_<\atop^{\sim}$}}
\newcommand{\gsim}   {\mbox{$_>\atop^{\sim}$}}
\newcommand{\lt}     {\mbox{$<$}}
\newcommand{\gt}     {\mbox{$>$}} 
\newcommand{\order}  {{\cal O}}
\newcommand{\JJ}     {{\cal J}}
\newcommand{\<}      {\langle}
\renewcommand{\>}    {\rangle}
\newcommand{\amin}   {\mbox{$^\prime\ $}}
\newcommand{\asec}   {\mbox{$^{\prime\prime}\ $}}
\newcommand{\ddeg}   {\mbox{${\rlap.}^\circ$}}
\newcommand{\beq}    {\begin{equation}}
\newcommand{\eeq}    {\end{equation}}
\newcommand{\beqa}   {\begin{eqnarray}}
\newcommand{\eeqa}   {\end{eqnarray}}
\newcommand{\ylm}[2] {{Y_{#1#2}}}
\newcommand{\ylmcc}[2]{{Y^{*}_{#1#2}}}
\newcommand{\fnlKS}  {f_{NL}^{\rm local}}
\newcommand{\fnleq}  {f_{NL}^{\rm equil}} 

In March 2008 the Wilkinson Microwave Anisotropy 
Probe (\wmap\ ) science team presented results on  
the analysis of five 
years of \wmap\ data. \wmap\ is a NASA Medium-Class Explorer
that yields full-sky maps of the
temperature and polarization anisotropy in the cosmic microwave background 
(CMB) of unprecedented fidelity. Simultaneous measurements in five 
frequency bands between 
23 and 94 GHz facilitate the separation of the CMB signal from foreground 
emission arising both from our Galaxy and from extragalactic sources. 
The CMB angular power spectrum derived from these maps exhibits a
coherent acoustic peak structure which makes it possible to extract a wealth 
of information about the composition and history of the universe, as well 
as the processes that seeded the fluctuations. Readers of ``Matters 
of Gravity'' will appreciate that the superb agreement between
CMB theory and measurement is a significant test of the General Theory 
of Relativity.

The ``{\sl WMAP5}'' release comprised seven papers. 
Hinshaw \etal\ (2008) summarize 
the basic results and report on data processing, calibration,
mapmaking, and systematic 
error limits with an emphasis on the changes since the {\sl WMAP3} release.
Hill \etal\ (2008)
discuss improvements in the physical optics model of the \map\
telescope, and use the results to determine the \map\ beam response.
Gold \etal\ (2008) report on the modeling, understanding, and 
subtraction of the temperature and polarized foreground emission.
Wright \etal\ (2008) analyze extragalactic point sources and 
present an
updated source catalog, with new results on source variability. 
Nolta \etal\ (2008) give the temperature and polarization
angular power spectra from the maps.
Dunkley \etal\ (2008) give the
parameter estimation methodology, the cosmological parameters inferred
from the \wmap\ data alone, and a comparison between different
cosmological data sets. They also develop an independent analysis
of polarized foregrounds.  Komatsu \etal\ (2008) present the 
cosmological interpretation of \wmap\ alone and in combination with 
a host of other cosmological data sets.
All 5 year \map\ data are available through NASA's Legacy Archive 
for Microwave 
Background Data Analysis (LAMBDA) and the data products are described 
in detail in the \map\ Explanatory Supplement by Limon \etal\ (2008), 
also available on LAMBDA.

\begin{figure}[t]
\centerline{\psfig{file=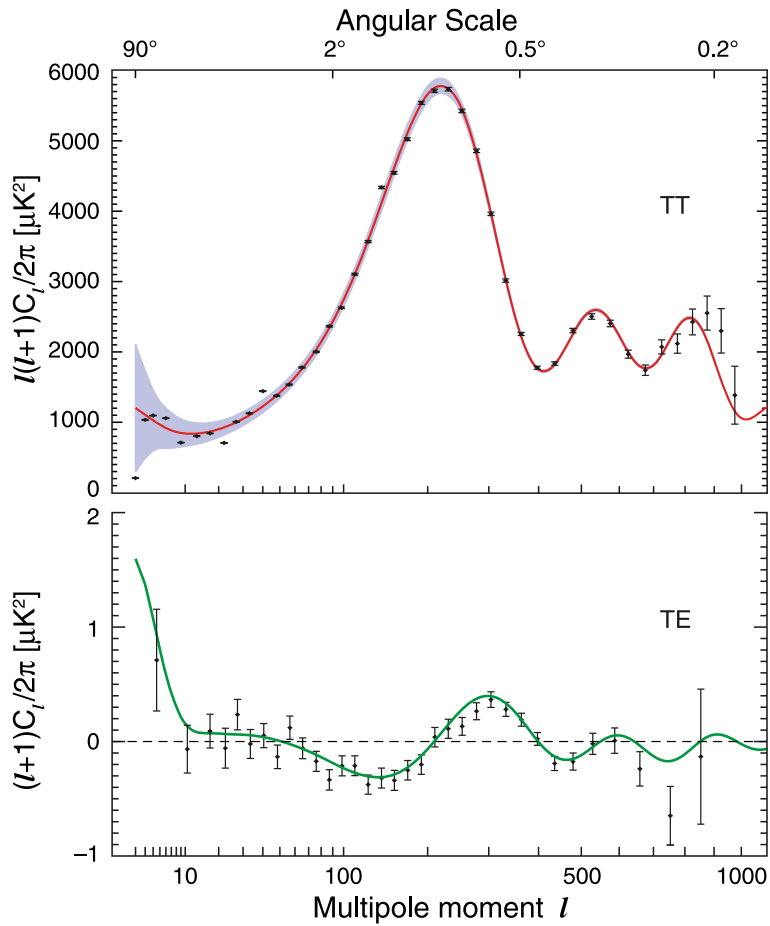,width=4in}}
\caption{The temperature (TT) and temperature-polarization correlation (TE)
power spectra based on the 5 year \map\ data.  The error bar per point
is shown. The line through the data is the best fit model. 
Cosmic variance is indicated by the grey band in the upper panel. The 
additional 2 years of data
provide more sensitive measurements of the third peak in TT and the high-$l$ TE
spectrum, especially the second trough. From Hinshaw \etal\ (2008).} 
\label{fig:tt_te}
\end{figure}

The primary data products from \wmap\ are the maps, their power spectra,
and an associated systematic error budget. 
To the eye, the 5-year maps look like the 3-year maps;
however, they have a much improved calibration and  
more detail at fine angular scales due to lower noise. In the limit 
that the CMB temperature 
fluctuations are normally distributed and that foreground emission has 
been properly subtracted over the full sky, the angular power spectrum
contains all the cosmological information. 
Figure~\ref{fig:tt_te} shows a version of the power spectrum binned in $\ell$.
The characteristic acoustic peak structure at decoupling, now seen by 
many different experiments (e.g., \cite{cmbexp}), is clearly evident in both 
the temperature (TT) and the temperature E-mode polarization cross 
correlation (TE). One model fits these substantially independent data sets. 

The ``standard cosmological model'' assumes a geometrically flat universe
composed of $\sim$4.6\% atoms, $\sim$23\% dark matter, and $\sim$72\% dark 
energy. The fluctuations are nearly scale invariant, adiabatic, and Gaussian.
It is stunning that all cosmological measurements, no matter the 
method or object under study, support this basic picture. Though we all seek
departures from the model, perhaps through non-Gaussianity or non-lambda-like
dark energy, an unambiguous departure has not been found.  In the standard
model, the oscillations in the primordial plasma that gave rise to the peaks
in Figure~\ref{fig:tt_te} were driven by a
gravitational landscape which was seeded at some very early time in the 
universe. The currently favored model for the source of the landscape is
quantum fluctuations expanded through an inflationary epoch.

For a good estimate of the cosmological parameters, one could simply fit
a model to the power spectrum in Figure~\ref{fig:tt_te} and adjust the 
parameters to minimize $\chi^2$. However, the precision of analysis demanded 
by the high quality of CMB data requires that the likelihood distribution 
at each $\ell$ for $\ell<32$, and the correlations between them, be taken 
into account.
This is done with codes that analyze together the maps, correlations, and power
spectra of the temperature and polarization. The results are reported in 
Dunkley \etal\ (2008) and Komatsu \etal\ (2008). 
The best fit model assuming a geometrically flat universe 
is given in Table~\ref{tab:summary} which was adapted from Komatsu \etal\ 
(2008).  The two  columns with results  give the mean values
for the \wmap\ data alone and \wmap\ in combination with baryon acoustic 
oscillation data (BAO, e.g., Eisenstein \etal\ (2005) , 
Percival \etal\ (2007)) 
and supernovae data 
(SN, e.g., Astier \etal\ (2006), Riess \etal\ (2007), Wood-Vasey \etal\ 
(2007)).

\begin{table}
\caption{Mean values and $1\sigma$ errors for parameters 
of the $\Lambda$CDM model.}
\begin{center}
\begin{tabular}{llcc}
Class
& Parameter
& \map\ 5-year
& \map5+BAO+SN\\
\hline
Primary &
$100\Omega_bh^2$
&\ensuremath{2.273\pm 0.062} 
&\ensuremath{2.265\pm 0.059} \\
&
$\Omega_ch^2$
&\ensuremath{0.1099\pm 0.0062} 
&\ensuremath{0.1143\pm 0.0034} \\
&
$\Omega_\Lambda$
&\ensuremath{0.742\pm 0.030} 
&\ensuremath{0.721\pm 0.015} \\
&
$n_s$
&\ensuremath{0.963^{+ 0.014}_{- 0.015}} 
&\ensuremath{0.960^{+ 0.014}_{- 0.013}} \\
&
$\Delta^2_{\cal R}(k_0)$ 
&\ensuremath{(2.41\pm 0.11)\times 10^{-9}} 
&\ensuremath{(2.457^{+ 0.092}_{- 0.093})\times 10^{-9}} \\
&
$\tau$
&\ensuremath{0.087\pm 0.017} 
&\ensuremath{0.084\pm 0.016} \\
\hline
Derived &
$H_0$
&\ensuremath{71.9^{+ 2.6}_{- 2.7}\ \mbox{km/s/Mpc}} 
&\ensuremath{70.1\pm 1.3\ \mbox{km/s/Mpc}} \\
&
$\Omega_b$
&\ensuremath{0.0441\pm 0.0030} 
&\ensuremath{0.0462\pm 0.0015} \\
&
$\Omega_c$
&\ensuremath{0.214\pm 0.027} 
&\ensuremath{0.233\pm 0.013} \\
&
$\Omega_mh^2$
&\ensuremath{0.1326\pm 0.0063} 
&\ensuremath{0.1369\pm 0.0037} \\
&
$t_0$
&\ensuremath{13.69\pm 0.13\ \mbox{Gyr}} 
&\ensuremath{13.73\pm 0.12\ \mbox{Gyr}} 
\label{tab:summary}
\end{tabular}
\end{center}
\end{table}

We first elaborate on the entries in the table, emphasizing
the last column, and then move on to other aspects of the model.
The emphasis follows that in Hinshaw \etal\ (2008).

\paragraph{Baryon density, $\Omega_bh^2$.} We have a precise (3\%) 
determination 
of the density of atoms in the universe. The agreement between the atomic 
density derived from \map\ and the density inferred
from the deuterium abundance is an important test of the standard big bang
model.  

\paragraph{Dark matter density, $\Omega_ch^2$.} We have a precise (3\%) 
determination of the dark matter density.  With five
years of data and a better determination of our beam response, this measurement
has improved significantly.  Previous CMB measurements have shown that 
the dark
matter must be non-baryonic and interact only weakly with atoms and radiation. 
The \map\ measurement of the density puts important constraints on
supersymmetric dark matter models and on the properties of other dark matter
candidates. The total matter density is given by $\Omega_m=\Omega_c+\Omega_b$
where we have ignored a negligible contribution from neutrinos.

\paragraph{Cosmological constant, $\Omega_\Lambda$.} In model with no 
spatial curvature (flat) and negligible radiation density, 
$\Omega_\Lambda = 1-\Omega_m$. This constraint is used to derive the 
Hubble constant. However, with the additional of BAO and SN data, we can relax
the flatness ($\Omega_k\ne0$) and the equation of state 
($w\ne-1$) constraints as done in
Komatsu \etal\ (2008). In the $\Omega_k-w$ plane, the WMAP+BAO 
and WMAP+SN lines
intersect at $\Omega_k\sim 0$ and $w\sim -1$. The combined constraints are
\ensuremath{-0.0175<\Omega_k<0.0085\ \mbox{(95\% CL)}}
and \ensuremath{-0.11<1+w<0.14\ \mbox{(95\% CL)}} further supporting the 
standard big bang model.

\paragraph{Early universe and scalar fluctuations.} 
\map\'s measurement of the primordial power spectrum of matter
fluctuations constrains the physics of inflation, our best model for the origin
of these fluctuations.  The primordial power spectrum is 
$P_{\cal R}(k)\propto k^{n_s}$. The 5 year data provide the best
measurement to date of the scalar spectrum's amplitude and slope ($n_s$).
With our definitions (Dunkley \etal, 2008), the amplitude is 
$\Delta^2_{\cal R}(k_0)=k_0^3P_{\cal R}(k_0)/(2\pi^2)$ where we
normalization at $k_0=0.002~{\rm  Mpc}^{-1}$.
It is noteworthy that without a model of the early universe, we would not 
have the prediction that $n_s$ should be slightly below unity. There is
now a strong connection between physical theories of the $10^{-35}$ second
universe and observations.
It should be noted that these constraints assume a smooth function of scale,
$k$.  Certain models with localized structure in $P(k)$, and hence additional
parameters, are not ruled out, but neither are they
required by the data.

\paragraph{Reionization, the optical depth $\tau$, and the birth of stars.} 
The first stars formed and reionized the universe with an optical
depth $\tau$. If the universe was reionized instantaneously starting from the 
neutral state we find $z_{\rm reion}=11.0$. In other words, the first stars 
took more than a half-billion years to turn on, yet still the universe was
reionized and the ``dark ages'' ended before the epoch of the oldest 
known quasars. We find $\tau$ through the large-angular-scale polarization 
measurements.
 
\paragraph{Hubble constant.} We have a precise determination of the 
Hubble constant, in conjunction with BAO
observations.  Even when allowing curvature ($\Omega_k \ne 0$) and a free dark
energy equation of state ($w \ne -1$), the data determine the Hubble
constant to within 3\%.  The measured value is in excellent agreement with
independent results from the Hubble Key Project (Freedman \etal, 2001),
providing yet another important consistency test for the standard model.

\paragraph{Age of the universe.} Because of the relation of the matter 
density, Hubble constant, and the manifestation of the acoustic
signature on the surface of last scattering, the CMB
can tell us the age of the universe. To percent level accuracy,
the universe is \ensuremath{13.7~\mbox{Gyr}} old. 

We now turn to addition tests and aspects of the basic model.

\paragraph{Adiabaticity of fluctuations.} We find significant constraints 
of the basic properties of the primordial
fluctuations.  The anti-correlation seen in the temperature/polarization (TE)
correlation spectrum on 4$^{\circ}$ scales implies that the fluctuations are
primarily adiabatic and rule out defect models and isocurvature models as the
primary source of fluctuations (Peiris \etal, 2003).

\paragraph{Neutrinos.} We see evidence for the presence of neutrinos in the 
early universe. Strictly speaking, as discussed in Dunkley \etal\ (2008),
we constrain any particle that is 
relativistic at decoupling ($z=1090$), couples very weakly to the 
baryon-electron-photon fluid, and has weak self-interactions. We associate 
neutrinos with these particles. At decoupling, some 380,000 years after the 
Big-Bang, neutrinos made up 10\% of the universe, 
atoms 12\%, dark matter 63\%, photons 15\%, and the dark energy was negligible.
The cosmic neutrinos were relativistic and existed in such huge numbers that 
they affected the universe's evolution. That, in turn, influenced the CMB. 
We are not able to tell the neutrino mass but constrain the sum of the neutrino
masses to be less than 0.61~eV (95\%CL).

\paragraph{Gaussianity.}
The statistical properties of the CMB fluctuations measured by \map\ are close
to Gaussian; however, there are several hints of possible deviations from
Gaussianity. Significant deviations would be a very important
signature of new physics in the early universe. In fact, searches for
non-Gaussianity may be the best avenue for identifying departures from the 
standard model.   The \wmap\ team has investigated a number of forms 
of non-Gaussianity, including the Minkowsky functionals, the one-point 
distribution, and the bispectrum,
and has not identified any clear signature of non-Gaussianity. However,
not all claims have been investigated.  

\paragraph{Primordial Gravitational Waves.}
In the original models of inflation, the ratio of the variance in primordial
fluctuations due to tensors (gravitational waves) to that of scalars (density 
fluctuations) was thought to be $r\approx0.2$, where $r$ is called the tensor
to scalar ratio.  The clearest way to identify the tensor contribution
may be through the polarization B-modes. \wmap\ does not yet have the 
sensitivity to do this. Instead, \wmap\ limits $r$ by
determining the largest possible contribution that tensors can make to
the temperature anisotropy. With a combination of BAO, SN, and \wmap5 data,
Komatsu \etal\ (2008) find $r<0.2$ at the 95\% CL. Some popular 
models of inflation are ruled out at $>3\sigma$.

\bigskip
\bigskip

There are many other aspects of the \wmap\ analysis---cosmological,
astrophysical  and instrumental---that are addressed in the papers.
What does not come across in a summary such as that presented 
here is the detailed analysis of the instrument that 
goes into making the science possible. Indeed
assessing systematic effects and testing the associated algorithms consumes 
by far the majority of the analysis effort. \wmap\ is still
fully functional and collecting data.

\vfill\eject

\newcommand{\gws}{gravitational waves~}
\newcommand{\gw}{gravitational wave~}
\section*{\centerline
{Science with LIGO}}
\addcontentsline{toc}{subsubsection}{
\it Science with LIGO, by Maria Alessandra Papa}
\parskip=3pt
\begin{center}
Maria Alessandra Papa, Albert Einstein Institute 
\htmladdnormallink{papa-at-aei.mpg.de}
{mailto:papa@aei.mpg.de}
\end{center}

The sensitivity band of LIGO extends between 50 Hz and 1500 Hz. In this band we expect gravitational wave signals from compact binary systems during their inspiral, coalescence and merger phases and from the oscillations of the object that forms after the merger. We expect \gws to be emitted during supernova collapse events; we also expect emission of continuous gravitational waves and a stochastic gravitational wave background. LIGO data is searched for all these types of signals.

Binary systems of compact objects evolve in orbits that gradually shrink in time, due to the emission of gravitational radiation 
.  The exact time-frequency evolution of the signal depends on a number of parameters, but the large timescales are set by the total mass of the system and systems with masses up to 200 solar masses are expected to emit signals with significant energy content in LIGO's band.  The sensitivity of a search for binary inspiral signals may be characterized by its horizon distance $d_H$. This is the distance at which an optimally located and oriented equal mass binary system is expected to produce a signal with matched-filter SNR = 8. 
Fig.\ref{fig:horizonS5} 
shows estimates of the horizon distance during the S5 run (the most sensitive and longest science run of LIGO \cite{whitcomb,lscExp07}), which
\addtocounter{figure}{-1}

\begin{figure}[!htbp]
\begin{center}
\includegraphics[width=0.8\textwidth]{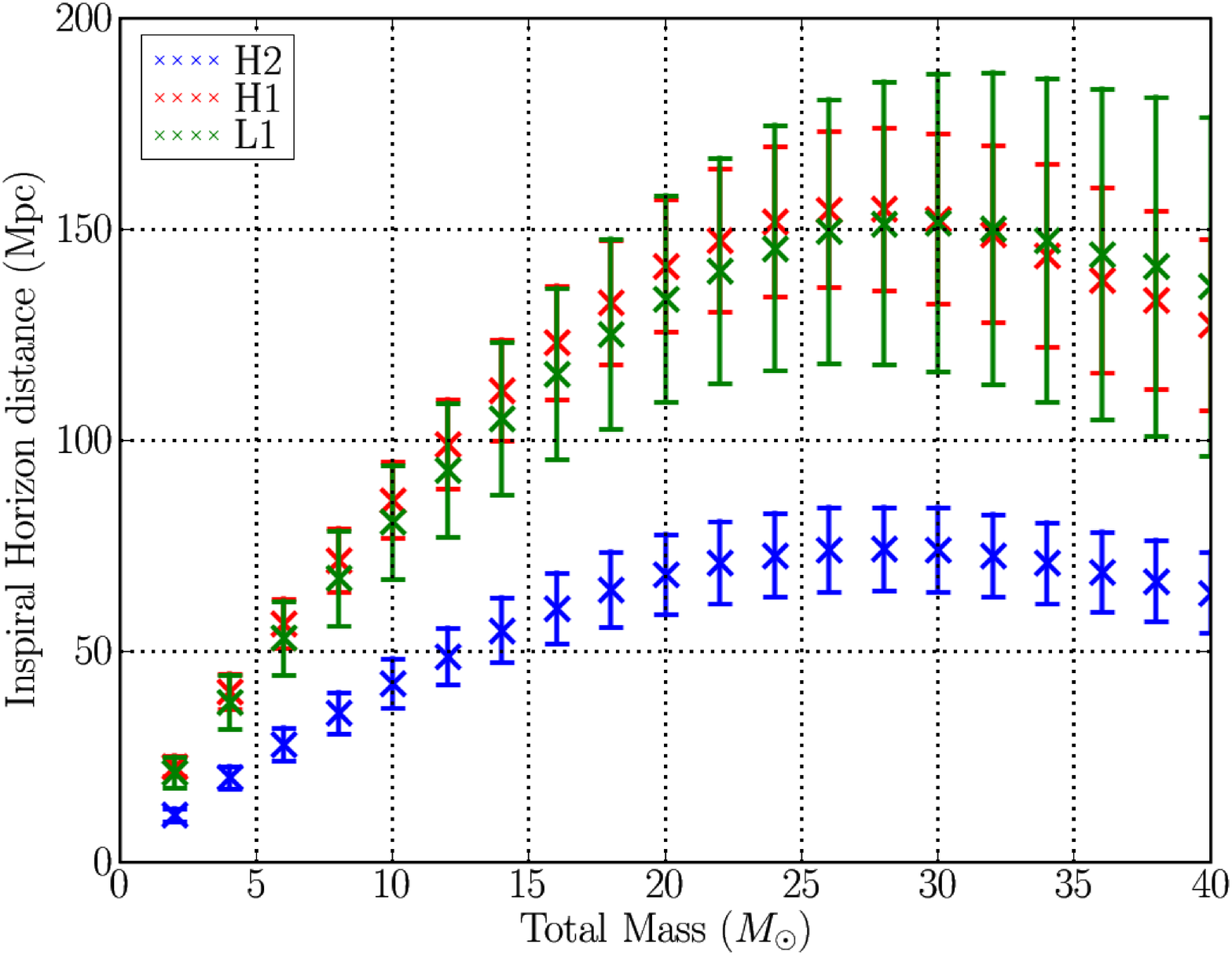}
\caption{Typical horizon distance during the S5 run as a function of the total mass of the binary system. H1 and H2 are the two Hanford 4 and 2 km baseline detectors. L1 is the 4 km detector in Louisiana. {\it Courtesy of the LSC}.}
\label{fig:horizonS5}
\end{center}
\end{figure}
comprise hundreds of Galaxies even for relatively low mass systems \cite{catalogue}. Still, the detection of a \gw signal from a binary inspiral signal is not at all ensured in S5 data: 
the S5 expected detection rates are 1 event per 400 to 25 years for 1.4-1.4 solar mass systems; 1 event every 2700 to 20 years for 5-5 solar mass systems and 1 event every 450 to 3 years for 10-10 solar mass systems. Enhanced detectors are expected to achieve an improvement in strain sensitivity of a factor of $\approx$ 2. With a horizon distance of 60 Mpc to neutron star systems the expected rates grow to 1 event every 60 to 4 years of actual observing time. Advanced detectors operating at a horizon distance of 450 Mpc to neutron star systems, bring the expected detection rates between several to order hundred events per year of observing time.

More sensitive than blind searches are triggered searches that take place when an independent observation is available. In February 2007 a short hard GRB, GRB 070201, was detected and localized within an area which includes one of the spiral arms of the M31 Galaxy. Since GRBs may be produced in the merger phase of binary neutron star systems (BNS) or neutron star-black hole binaries (NSBH) this particular GRB could well have been associated with a detectable \gw signal if coming from M31, at 800 kpc. An inspiral search was carried out on the available \gw data for systems with component masses in the range 1-3 and 1-40 solar masses respectively but no signal was found \cite{grb070201}.  
This null result excluded the possibility that the GRB be due to a binary neutron star or NSBH inspiral signal in M31 with very high confidence (greater than 99\%). It also excluded various companion mass - distance ranges significantly further than M31, as shown in Fig. 3 of \cite{grb070201}.

A search for a burst signal associated with GRB070201 was also carried out and resulted in an upper limit on the isotropic \gw energy emission at the distance of M31 around $150$ Hz of $\approx 8\times 10^{50}$ erg. This result is significantly less informative than the one from the inspiral searches because the upper limit is orders of magnitude larger that the estimated energy release in gamma rays at the same distance. A soft gamma ray repeater (SGR) flare event in M31 is consistent with the gamma-ray energy release and is not ruled out by the \gw analysis \cite{ofek}.

A systematic search of \gw signals associated with 191 SGR bursts was carried out using S5 data and prior data in coincidence with the 27 Dec 2004 giant flare from SGR 1806-20 \cite{s5SGRs}. No signals were found and upper limits on the isotropic \gw energy were placed. At a nominal distance of 10 kpc these upper limits overlap with the range of electro-magnetic isotropic energy emission in SGR giant flares ($10^{44}$ - $10^{46}$ erg) and some of the upper limits on the ratio of the \gw and electromagnetic energies are within the range of theoretically possible values.

In general there are many circumstances in which short bursts of \gws are expected, lasting from a few ms to a few seconds involving the merger phase of a binary system or the collapse of a stellar core. Blind searches for these types of events are routinely carried out. 
Preliminary estimates of the reach of S5 burst searches show that, at best, 50\% detection efficiency can be achieved for signals generated by converting of order 5\% of a solar mass at the distance of the Virgo cluster, or $\sim 2\times 10^{-8}$ of a solar mass at the Galactic center. Estimates of the expected amplitude of burst signals vary quite widely and scenarios exist which predict emission that is detectable in S5. For example \cite{baker06} predicts for black hole mergers the emission of up to 3\% of solar masses in gravitational waves. A system of this type formed by two 50 solar mass black holes at $\approx$ 100 Mpc would produce \gws which could be detected with 50\% efficiency in S5.


An isotropic stochastic background of gravitational radiation is expected due to the superposition of many unresolved signals, both of cosmological and astrophysical origin. The background is described by a function $\Omega_{GW}(f)$, which is proportional to the energy density in \gws per logarithmic frequency interval. The most recent results from searches for isotropic backgrounds come from the analysis of the S4 LIGO data and for a flat \gw spectrum put a 90\% Bayesian upper limit at $\Omega_{GW} \times \left[ {H_0\over {72 {\rm {km s^{-1} Mpc^{-1}}}}}\right] < 6.5\times 10^{-5}$, in the frequency range 50-150 Hz \cite{S4stoch}. This limit is still above the one that may be inferred from measurements of light-element abundances, WMAP data and the big bang nucleosynthesis model, but it is expected that the data from the S5 run will probe values of $\Omega_{GW}$ below this.

Fast rotating neutron stars are expected to emit a continuous \gw signal if they present a deviation from a perfectly axisymmetric shape, if their r-modes are excited, or if their rotation axis is not aligned with their symmetry axis (\cite{S2Fstat}). In all cases the expected signal at any given time is orders of magnitude smaller than any of the short-lived signals that have been described above. However, since the signal is present for a very long time (to all practical purposes, in most cases, one may consider it there {\em{all}} the time), one can increase the SNR by integrating for a suitably long time. \
 
No gravitational wave signal has been detected while searching for continuous \gws from known radio pulsars. This is not unexpected because for most systems the indirect upper limit on the amplitude of \gws that one may infer from the measured spin-down rate of the systems is more constraining that the limit determined by the \gw observations. However in one case \gw observations are actually beating the electromagnetic spin-down limit and starting to probe new ground. This is the case of the Crab pulsar. With 9 months of S5 data LIGO observations beat the spin-down upper limit by a factor of about 4 \cite{Crab}. More importantly, assuming phase coherence between the \gw and radio signals, the \gw luminosity is constrained to less than 6\% of the observed spin-down luminosity. On other pulsars, albeit not beating the spin-down upper limits, the LIGO results are expected to reach values as low as a few $10^{-26}$ in the intrinsic gravitational wave amplitude $h_0$ \cite{jks} and several $10^{-8}$ in ellipticity $\epsilon$. These results show that at the current sensitivity, it is possible that LIGO could detect a continuous \gw signal, coming from an unusually nearby object, unknown electromagnetically and rotating close to $\sim 75$ Hz.

\begin{figure}[!htbp]
\begin{center}
\includegraphics[width=0.95\textwidth]{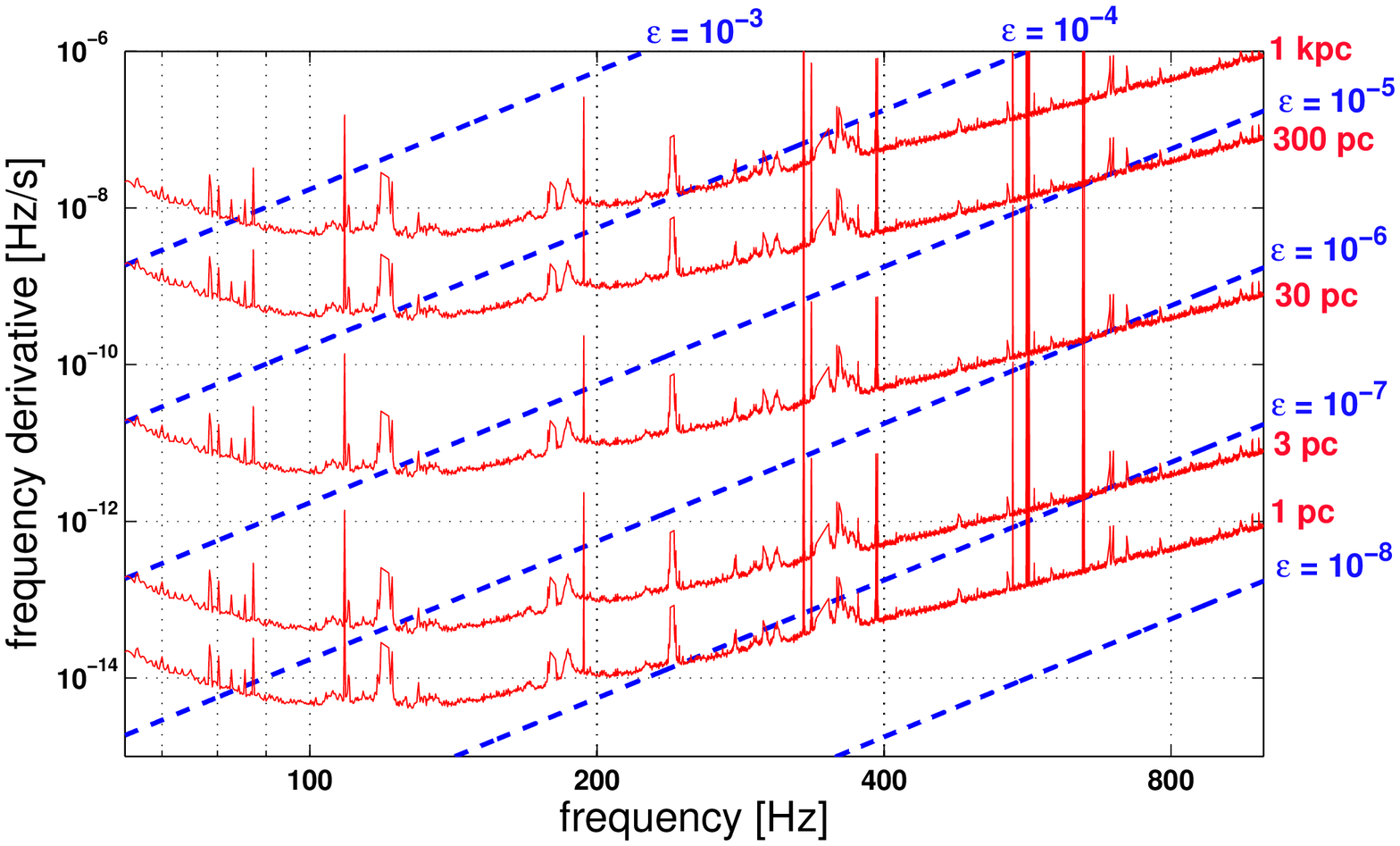}
\caption{Solid curves: Frequency and frequency-derivative values of a signal that would be detectable by the S4 stack-slide search described in \cite{S4PHS}. Dashed lines: lines of constant ellipticity.}
\nonumber
\label{fig:blindSearchReach}
\end{center}
\end{figure}

The most promising searches look for previously unknown objects, and are often refered to as blind searches (\cite{S2Fstat,S2Hough,S4PHS,S4EatH}). 
Deep blind searches require an enormous amount of computational power and in fact are carried out by Einstein@Home, a public distributed computing project that uses compute cycles donated by the general public. Einstein@Home is the second largest public compute project in the world and delivers an average 100Tflops of compute power continuously \cite{E@h}.
In the absence of a detection, upper limits are placed on the intrinsic amplitude of the \gw signal at the detector, $h_0$, and 
may be recast as frequency-first frequency derivative curves which represent an excluded \gw signal at a fixed distance. On the same plane one can overlay constant ellipticity $\epsilon$ curves and understand what ellipticity values the distance parametrized curves refer to. 
Fig.\ref{fig:blindSearchReach} 
shows this type of curve, deduced from the Hanford 4km interferometer S4 data stack-slide search upper limits of \cite{S4PHS}. In S5, the most sensitive Einstein@Home searches are expected to yield a sensitivity improvement in $h_0$ close to a factor of 10, resulting in a detectability range of $\sim$ 1 kpc at $150$ Hz with $\epsilon \sim 10^{-5}$.

\section*{Acknowledgements}
I am grateful to many colleagues in the LIGO Scientific Collaboration for valuable discussions and in particular to Ray Frey for his useful comments.
This brief was assigned the following LIGO document number: P080096-00-Z

\vfill\eject

\section*{\centerline
{7th LISA Symposium}}
\addtocontents{toc}{\protect\medskip}
\addtocontents{toc}{\bf Conference reports:}
\addcontentsline{toc}{subsubsection}{
\it 7th LISA Symposium, by Edward K. Porter}
\parskip=3pt
\begin{center}
Edward K. Porter, Albert Einstein Institut
\htmladdnormallink{ed.porter-at-aei.mpg.de}
{mailto:ed.porter@aei.mpg.de}
\end{center}

The LISA Symposia occur every two years, with the location alternating between Europe and the U.S.  The $7^{th}$ LISA Symposium was held this year in Barcelona, Spain from the $16^{th}-20^{th}$ July.  The Symposium was jointly organized by the Institute of Space Sciences (ICE) [National Spanish Research Council (CSIC) \& Institut of Space Studies of Catalonia (IEEC)], and the local organizing committee was
the LISA group led by Alberto Lobo.  The venue for this year was the excellent science museum CosmoCaixa.  The Symposium was sponsored by
the European Space Agency, the Albert Einstein Institute, the La Caixa Foundation, the Spanish Ministry of Science and Education (MEC), the Catalan Agency for Research and University Funding (AGAUR), the National Spanish Research Council, the Spanish Society of Gravitation and Relativity (SEGRE), the University of Barcelona (UB), the Polytechnic University of Catalonia (UPC), and the Institute for High Energy Physics (IFAE). 

Each day started with a number of plenary lectures, splitting off into parallel sessions in the afternoons.  While the status of LISA has been enjoying a bit of a rollercoaster ride in the last few years, it was nice to see that the number of people working on LISA and LISA Pathfinder continues to grow.  This year's Symposium had 226 registered participants.  There were 30 plenary lectures, 78 parallel session presentations and 75 poster contributions.  The lists of speakers and sessions can be found on-line at

\htmladdnormallink
{http://www.ice.cat/research/LISA\_Symposium}
{http://www.ice.cat/research/LISA\_Symposium}\\

On the Wednesday of the meeting, there was an outreach lecture by Prof. 
Clifford Will entitled ``Black holes, waves of gravity and other warped 
ideas of Dr Einstein.''

The plenary talks covered a large area this year.  On the first day we heard 
about how LISA is fitting into both ESA's Cosmic Vision Plan for 2015-2025 and 
NASA's Physics of the Cosmos theme.  There were a number of talks about LISA and LISA Path Finder (LPF) technology, engineering advances for both projects, data analysis for LPF and the question of optimal mission size and duration.  In the weeks leading up to the Symposium, there had been a call from NASA to investigate mission downsizing for LISA.  The outcome of this investigation was that there would be a large loss of science for a small saving of money.  Therefore, it looks like the proposed mission is the optimal mission.  There was also a presentation about the status of the ground based detectors.  Other talks covered topics such as advances in data analysis for LISA and the growing link between data analysts and astrophysicists.  One highlight of the meeting was the advance in numerical simulations and numerical relativity.  We were treated to some very interesting presentations on structure formation at galactic centers and the numerical modelling of inspiral waveforms, through merger and into the ringdown phase.  One of the other motivating aspects is the advance in data analysis techniques since the last LISA Symposium.  We are now starting to investigate more difficult sources such as spinning black hole binaries, extreme mass ratio inspirals 
(EMRIs) and more realistic galaxy models.

The parallel sessions for LISA/LPF had a heavy emphasis on technology for LPF due to the rapidly approaching launch date.  One of the main topics of discussion was LPF data analysis and the types of science that could be done with LPF.  There were also presentations on the LISA Technology Package (LTP).  
The goal of this experiment is to investigate the sources of disturbances that may cause the test masses to be perturbed from geodesic trajectories.  There is work being done on the development of a data analysis algorithm, and there was also the announcement of the first LTP Mock Data Challenge.  Finally, there were presentations on other detectors such as DECIGO, Super-Astrod and GRACE, and how technology from both LISA and LPF could be used to enhance these missions.

The theory parallel sessions could normally be divided into three distinct areas : data analysis, astrophysics and numerical relativity.  One of the most uplifting aspects of this particular meeting was seeing that the divisions between the areas has really started to become blurred.  Groups from all three areas are 
now communicating and working with each other to develop search algorithms 
based on more realistic initial data and waveforms.  The main areas of interest for a lot of the presentations were the inspirals of massive black holes and EMRIs.  These focused on more realistic EMRI waveforms, binary inspirals with 
higher harmonic corrections, spin, eccentricity and numerical mergers.  
There were also discussions on the detection of Intermediate Mass Binaries, 
the possibility of detecting cosmic superstrings and constraining both 
string theory and dark energy predictions with LISA.

The Proceedings of the Symposium will be published jointly by Classical and 
Quantum Gravity and the Journal of Physics: Conference Series of the 
Institute of Physics (IoP). 

From the level of participation, and the advances that have been made since the last LISA Symposium, the community is looking very healthy and is on track for both the launches of LPF and LISA.

\vfill\eject

\section*{\centerline
{The Fourth Gulf Coast Gravity Conference 4(GC)$^2$}}
\addtocontents{toc}{\protect\medskip}
\addcontentsline{toc}{subsubsection}{
\it The Fourth Gulf Coast Gravity Conference 4(GC)$^2$, by Lior Burko}
\parskip=3pt
\begin{center}
Lior Burko, University of Alabama in Huntsville
\htmladdnormallink{burko-at-uah.edu}
{mailto:burko@uah.edu}
\end{center}

The Fourth Gulf Coast Gravity Conference was hosted by the University of Mississippi in Oxford, MS March 7 and 8, 2008. The 40 participants  witnessed some freak weather, as it was heavily snowing (well, heavy for Mississippi in March...) and the University closed down at the exact hour that the meeting was about to start. The hosts, Marco Cavagli\`a and Luca Bombelli, produced a well organized meeting with the generous funding of the Department of Physics at Ole Miss, and the author of this summary, as well as all the other participants, wishes to thank them. 

Twenty four talks were given, including ten by students who competed for the Topical Group on Gravitation's prize for the student who presented the best talk. 

Tyler Landis (LSU) gave the first talk, describing the satus of the multi--patch general relativistic magnetohydrodynamics code developed at LSU. Wolfgang Rindler (UTD) described his work on the contribution of a cosmological constant to the bending of light in a Schwarzschild--de Sitter universe, in the swiss cheese model, arguing that an older claim by Islam for a null effect was incorrect.  The effect found by Rindler could be observed with gravitational lensing. Brian Mazur (Ole Miss) discussed high-velocity cloud interactions with the glactic halo, and was followed by Gerrit Verschuur (U.~of Memphis) who talked about a possible correlation of the WMAP results and the galactic interstellar hydrogen features. Verschuur described evidence for such a scenario, suggesting that the actual cosmological structure at the time of recombination was much smaller than usually thought, arguing that the observed structure has important galactic contributions. The statistical significance of these intriguing correlations was strongly debated by the participants.

Marco Cavagli\`a (Ole Miss) gave his interpretation of a Dickens classic, discussing the ghosts of LIGO's past, present, and future, in particular the analysis of a recent GRB and ruling out its hypothesized source in the Andromeda galaxy. Jun--Qi Guo (Ole Miss) discussed data quality vetoes for high--mass compact binary coalescences in LIGO's S5 run, and Myungkee Sung (LSU) discussed the optimal filter technique for detection of gravitational waves with LIGO. Lior Burko (UAHuntsville) discussed a Monte Carlo approach for the calculation of black hole quasi--normal modes as an alternative for the current all--sky average approach, and its application for LISA. Most importantly, Burko argued that for Kerr black holes the many source limit of the Monte Carlo approach does not coincide with the all--sky average results. Steve Detweiler (U.~of Florida) reminded the participants of the early days of quasi--normal modes research in the 1970's, and the various names that were in use at the time, including his semi--serious proposal to name them after a human bodily function that reflects their poor quality factor. Steve then continued with a discussion of self force regularization incurve spacetime making use of the singular field.

Saturday talks started with Peter Diener (LSU), who demonstrated his adoption of Louisiana culinary tradition by discussing his turducken recipe. Peter emphasized his insight that  a well--stuffed turducken looks just like turkey. And no, Peter did not actually treat us to a turducken dinner, a failure he will have a second chance to remedy when the Fifth Gulf Coast Gravity Conference is hosted next year at LSU. A menu of turducken served with Peter's home baked Danish sourdough beer bread would surely attract many to this important meeting. 

Oleg Korobkin (LSU) discussed a finite element approach for solving  constraint equations on multi--block triangulations, and Frank Loeffler (LSU) described numerical codes for mixed binaries of a black hole and a neutron star.  Pedro Marronetti (FAU) discussed high-spin binary black hole mergers, and the highest spin likely to be created by nature.  Ian Vega (U.~of Florida) discussed the application of the self force regularization approach to circular Schwarzschild orbits for a scalar toy model in the time domain. Agreement with high accuracy frequency domain results is impressive. Paul Walter (UTA) described the status of OpenGR, an open framework for doing large general relativistic simulations.

Sergio Fabi (UA) discussed the noncommutative geometry approach to zero point energy on extra dimensions. Fabio Scardigli (Kyoto U.) described properties of micro black holes and their possible creation in the LHC, and Usama al--Binni (U.~of Tennessee Knoxville) discussed black holes on the brane with tension. James Alsup (U.~of Tennessee Knoxville) described Bjorken flow from an AdS-Schwarzschild black hole, and Brett Bolen (Western Kentucky U.) discussed the motivation underlying the quest for having a minimal length scale. Luca Bombelli (Ole Miss) discussed dynamics of causal sets, and Arunava Roy (Ole Miss) addressed discriminating SUSY and black hole at the LHC. Hristu Culetu (Ovidius U., Romania) discussed the Doran--Lobo--Crawford time dependent spacetime, and Alan Stern (UA) discussed discrete spectra from noncommutative geometry, including a quantization of the cosmological constant in a noncommutative Chern--Simons theory. 

And the Topical Group on Gravitation Prize for the student who gave the best talk at the Fourth Gulf Coast Gravity Conference, including an actual blue apple trophy, went to ... Ian Vega. Congratulations Ian!

The meeting's website includes abstracts of all talks, at the following URL: 

\htmladdnormallink
{http://www.phy.olemiss.edu/GR/gcgm4/index.html}
{http://www.phy.olemiss.edu/GR/gcgm4/index.html}\\

\end{document}